# Topological measures for the analysis of wireless sensor networks

Vincent Labatut[a*], Atay Ozgovde[b]

[a]*CompNet Group, Galatasaray University, Istanbul, Turkey*
[b]*Pera Lab, Galatasaray University, Istanbul, Turkey*

**Abstract**

Concepts such as energy dependence, random deployment, dynamic topological update, self-organization, varying large number of nodes are among many factors that make WSNs a type of complex system. However, when analyzing WSNs properties using complex network tools, classical topological measures must be considered with care as they might not be applicable in their original form. In this work, we focus on the topological measures frequently used in the related field of Internet topological analysis. We illustrate their applicability to the WSNs domain through simulation experiments. In the cases when the classic metrics turn out to be incompatible, we propose some alternative measures and discuss them based on the WSNs characteristics.

*Keywords:* Wireless Sensor Networks; Complex Networks; Topological Measures: Sink-Distance; Sink-Betweenness

## 1. Introduction

Wireless Sensor Networks (WSNs) are composed of resource-limited, small form factor sensor nodes spatially distributed over a region of interest for monitoring purposes. Their use spreads across many domains such as military (border surveillance), environmental (forest fire detection), industrial (machine health monitoring) and medical (clinical monitoring) applications. Such a network contains at least one sink node, which is used as a gateway towards a server or an external network such as the Internet. In a typical application, the sensor nodes are self-organized to collaboratively capture and report data to the sink nodes. Although categorically being a communications network, WSNs with their intrinsic properties may present radically different characteristics. For instance, the designated sink node is the ultimate recipient for all the data flow in the network. This creates a many-to-one type (convergcast) traffic which is more difficult to handle than a many-to-many traffic matrix. Another atypical aspect is due to the finite and unreplenishable initial energy deposited in the sensors. This fact causes WSNs to be designed using algorithms and approaches that gives strict priority to energy efficiency. Unique properties of WSNs have been studied, especially under the communication networking domain and many novel routing, MAC algorithms specific to WSNs have been devised. In this work, to further understand the properties of the WSNs, we explore them using the tools of complex networks research domain.

[*] Corresponding author. Tel.: +90 (0)212-227-44-80 / 457; fax: +90 (0)212-259-20-85.
*E-mail address*: vlabatut@gsu.edu.tr.

Complex networks are a widespread tool for the modeling of complex systems, they are used in many domains, including biology, communication, sociology, physics, economy, etc. [1]. In the recent years, a whole arsenal of measures was defined to study them. More interestingly for us, they have been intensively used to study computer networks, especially the Internet. Various methods were used to retrieve the Internet structure, focusing on different granularities such as the inter-domain [2-5] and router [4, 5] levels. Using such a topological approach to study computer networks leads to a better understanding of the network structure, which in turn brings multiple benefits. First, improving the models used to provide simulation data, resulting in more realistic testing conditions [3, 5, 6]. All the tools specifically developed for the network can therefore be assessed more accurately. Moreover, it is also possible to consider the network structure from a dynamic perspective. By studying the evolution of the topology, one can derive predictive models [7]. A better understanding of the network structure also allows designing better management methods and tools [6]. Finally, by taking advantage of this knowledge, tone can improve communication protocols by adapting them to the network structure [2, 3].

Stemming from the fact that WSNs belong to the class of computer networks, such a topological approach seems relevant to improve the tools and technologies related to them. Several works already applied complex network research tools to WSNs to better understand their dynamics and enhance their performance [8, 9]. Moreover, in the case of WSNs, topological measures are particularly interesting because they allow comparing one network at different time steps, or several distinct networks. This makes possible assessing the effect of the deployment (spatial distribution, number of sinks, etc.) and operational (communication protocol, type of sensor, etc.) parameters on WSN-specific features, such as the network life time. However, from the topological point of view, WSNs may present considerable differences from mainstream computer networks, particularly the Internet: their temporal scale is much smaller, they shrink instead of growing, and they are structured around certain nodes with specific properties, i.e. the sinks. For this reason, topological measures for the assessment of WSNs have to be chosen with care, as the classical measures may not be directly applicable. In this article, we try to answer this question by reviewing the main measures used in previous works to study the Internet, and discussing their suitability to the analysis of WSNs. For the cases where the studied measure turns out to be inappropriate, we define alternatives and explain the intuition behind them. In order to illustrate our arguments, we use some simulated WSN data generated thanks to Opnet.

The rest of the article is organized as follows. In section 2, we describe the method we followed to simulate a WSN, and generate the data later used as an example when presenting the measures. Each following section (3-6) then corresponds to a family of classic topological measures: density-, degree-, distance- and centrality-based. For each family, we first present the classic measures and how they are interpreted when considering the Internet. We then discuss their relevance and interpretation in the context of WSNs. If they are not suitable, we define some more appropriate versions and describe their properties and use. In both cases, we use our simulation data to show the type of results obtained through the measure. Finally, in the very last section (7), we discuss our work and contributions, and propose some extensions.

## 2. Description of the simulation

We have employed *OPNET Modeler* to perform packet level simulations where the individual sensor behavior, network interactions and energy consumption are modeled realistically. The monitored region is a $100 \times 100$ m square containing $100$ *sensors* and a single *sink*. The sensors are uniformly randomly deployed, whereas the sink is statically located at the geographical center. The sensing radius is 10 m, and the initial sensor energy is 1 Joule. Their data communication rate is of 20 kbps. In order to construct the base communication network, we executed a setup phase by applying the distributed Bellman-Ford algorithm: sensors broadcast their energy-wise distance to the sink. To enhance energy efficiency, sensors are allowed to use 4 discrete power levels for RF transmission, which they can dynamically adjust depending on the specific next hop. The communication range is at least 10 m and at most 50 m.

The routing algorithm used in the experiments is the *Basic Probabilistic Routing* (BPR) in which sensors operate on local information about the residual energy levels of their neighbors. According to BPR, the probability of a neighbor being chosen as the next hop is inversely proportional to its distance (energy cost) from the sink. This is in contrast with Minimum Energy Routing (MER), where the sensors greedily choose the least cost paths to the sink, which results in heavy usage of certain sensors. BPR, thus, tends to favor least cost paths but also balances the traffic load by letting sensors occasionally chose sub-optimal paths, which results in increased network lifetime. We set a limit of 3 neighbors top for each sensor. In our simulations, to further enhance the lifetime, sensor operations are categorized under NORMAL and SELFISH modes. This mode directly determines the routing behavior of the sensors. When the normalized residual energy of a sensor drops below a predefined threshold of 0.05, the sensor switches from NORMAL to SELFISH mode and broadcast this change to the rest of the network with maximum transmission power. In the SELFISH mode, sensors only transmit their own sensed data and do not function as relay sensors.

Our goal with these example data is to illustrate how one can use the presented measures (both classic and new ones), and interpret them in order to characterize the topology of a WSN.

## 3. Nodes, links and density

**Nodes and links.** The most basic properties to characterize a network are the numbers of *nodes* $n$ and *links* $m$ it contains. They allow studying how the size of a given system evolves. For instance, they were used to monitor the growth of the Internet through the comparison of several snapshots [2-4, 10]. They are particularly interesting for the WSNs, as a tool to quantify how such networks shrink. In this context, we propose to distinguish the *isolates* and the other nodes, whose counts we note $n_-$ and $n_+$, respectively. The former do not have any connection and therefore correspond to dead (or disconnected) sensors, whereas the latter represent the sensors still alive and connected to the sink. They allow monitoring the part of the WSN which is no longer functioning, as shown in Fig. 1-a. Any sudden change in $n_-$ suggests a large part of the network just died at once. In our experiment, the proportion of dead sensors increases regularly, almost linearly with time, so we can infer only a few nodes are disconnected at each time step.

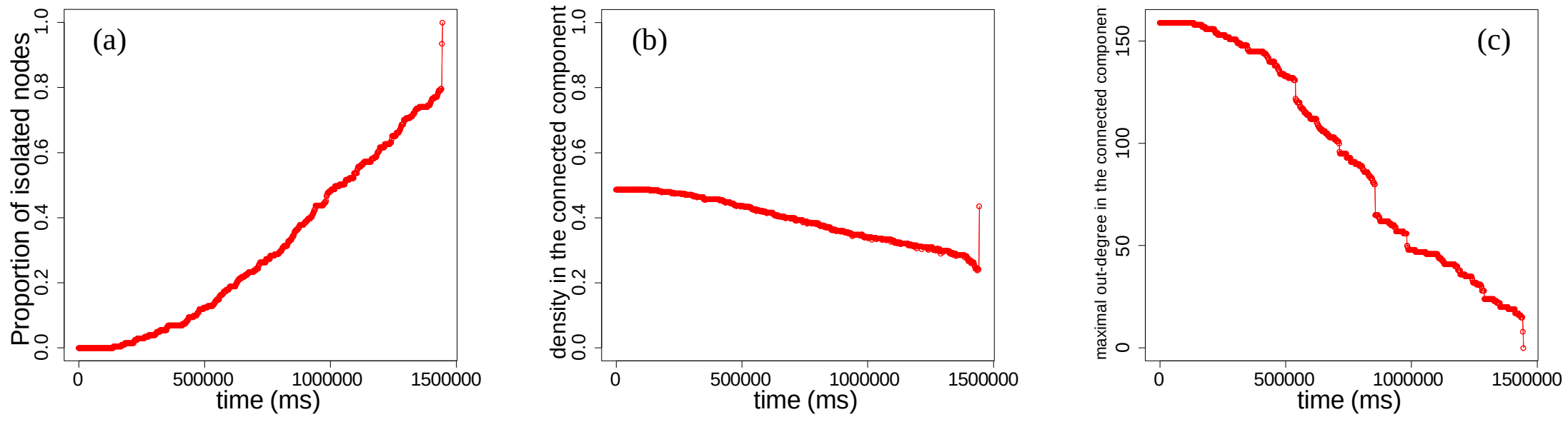

Fig. 1. (a) proportion of isolates $n_-/n$, (b) density in the connected component $d_+$ and (c) maximal out-degree, in function of time.

**Density.** The *density* $d$ of a network is directly related to its numbers of nodes and links: it corresponds to the ratio of the numbers of existing to possible links. The latter is derived by considering a fully connected network containing the same number of nodes. Consequently, if the network is directed, as it is our case, then $d = m/n(n-1)$. The density ranges from 0 (no link at all) to 1 (fully connected network). In the case of the WSNs, the overall density is largely influenced by the increasing proportion of isolates. Since these are already handled by $n_-$, we propose to focus only on the connected component (sensors still able to reach the sink) with $d_+ = m/n_+(n_+ - 1)$. Any significant variation in $d_+$ suggests that the disconnection of one or several sensors caused an important modification of the connected component, or alternatively that many links disappeared at once. Note this variation can be positive or negative, depending on the relative density of the disconnected part. As shown in Fig. 1-b, the density of the network in our experiment is initially very high (when compared to other real-world networks), and then regularly decreases with time. This decrease undergoes a short acceleration at the end of the simulation; however even the density observed just before the death of the network remains very high. The observed regularity means the amount of links removed at each time step is stable.

## 4. Degree-based measures

**Overall degrees.** The *degree* of a node (also called connectivity in the context of computer networks [5]) corresponds to the number of links connected to it. In the case of a directed network, one can distinguish the *in*- and *out-degrees*, i.e. the counts of incoming and outgoing links, respectively. Their sum is called the *all-degree*. In the case of the Internet, many authors studied some statistics related to the degree, such as the mean [3, 10] and maximum [4]. The *average* all-degree, in particular, is sometimes used as an alternative to the density for assessing the sparsity of the network [2]. For WSNs, the minimal and maximal degrees are particularly revealing. Indeed, a significant change in the *maximal* degree means either the most connected nodes lost many links, or were all disconnected, at once. If such a node is unique, it generally holds an important role in the network. If there are several ones, then such a change is likely to be caused by a non-trivial modification of the network structure. In both cases, identifying the time step of this change is of great interest.

For our experiment, the maximal out-degree plot displays some clearly visible changes (Fig. 2-c), corresponding to the death of highly connected nodes. The *minimal* degree can also bring some relevant information, provided one focuses on the connected component. For instance, in our experiment, sensors can enter the selfish mode, therefore removing all their incoming connections. Such moments appear as punctual 0 values. Both these measures should be considered as complementary to those presented in the previous section, because the changes monitored here do not necessarily reflect on these. Indeed, a link removal does not imply a node removal ($n_-$), and the modifications which concern the optima can be neglectable relatively to those undergone by the rest of the network ($d_+$).

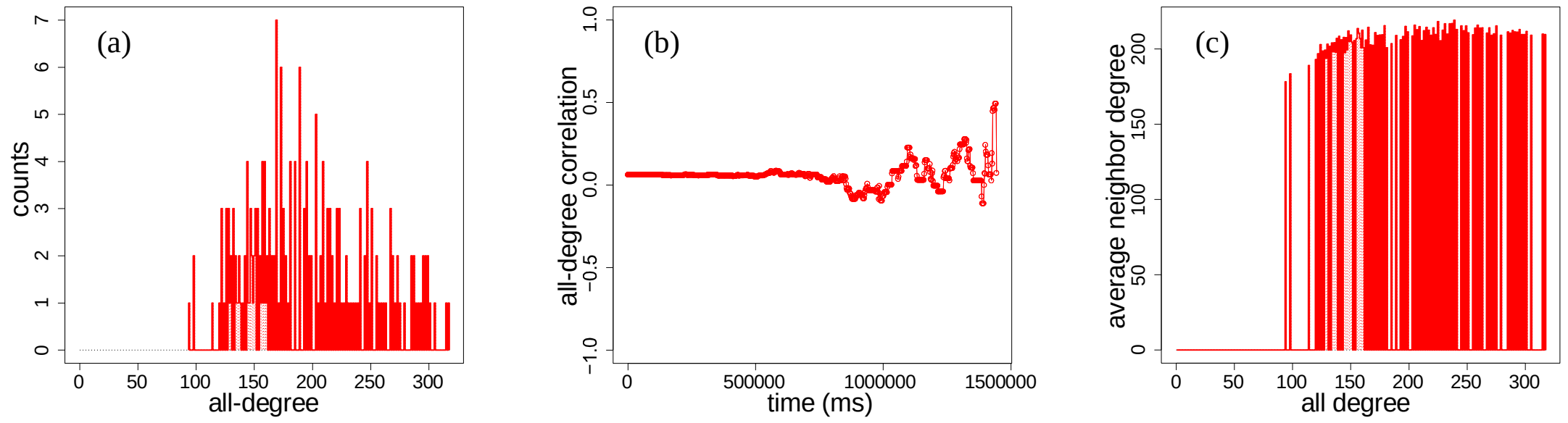

Fig. 2. (a) all-degree distribution in the connected component; (b) degree correlation in function of time; (c) average neighbour degree in function of the degree at $t = 1393200$.

**Degree distributions.** However, very different networks can have the same average, maximal or minimal degree, which is why some authors prefer to study the whole *degree distribution* [3], which is more informative. When studying the Internet, some authors [2] distinguished several parts in the degree distribution and used them to identify different categories of nodes, fulfilling different roles (e.g.: leaves, relays, centers…). Others identified the nature of this distribution (e.g. power-law) and summarized it using its estimated parameters (e.g. exponent) [4], in order to characterize or compare networks. For WSNs, the degree distribution is more difficult to analyze because of the temporal definition: one distribution can be generated at each time step. So one may want to focus on moments corresponding to significant changes in other properties (density, average degree, etc.). In our experiment, all degrees are normally distributed in the connected component (Fig. 2-a), and this remains true until the end of the simulation. This can be explained by the uniform spatial distribution of sensors: the more peripheral the sensor and the smaller its degree.

**Assortativity.** Besides the degree itself, the *assortativity* of the network constitutes another important property [11]. It indicates how much nodes tend to connect to other nodes with similar degree. The most straightforward and simple way to study it is to consider the degree-degree correlation $\rho$. A significant departure from zero indicates a strong hierarchical structure. However, while studying the internet, authors prefer to use a finer tool based on the *average neighbor degree* $\langle k_{nn} \rangle$ instead of the plain correlation. This value is obtained for a given node by averaging the degrees of its direct neighbors. One generally considers it expressed as a function of the degree of the considered node [5, 10], and notes it $\langle k_{nn} \rangle_k$. It allows not only detecting the presence of a hierarchical structure, but also charactering its nature.

In the context of WSNs however, the degree correlation constitutes a relevant measure because of the very dynamic nature of the system. It allows identifying important changes in the network structure, and then focus on the concerned time steps by studying $\langle k_{nn} \rangle_k$ for these specific moments. As shown in Fig. 2-b, the network from our experiment is clearly not assortative, with a correlation remaining very close to zero until the network starts getting very small. Again, this can be explained by the spatial distribution of the sensors. Because of the physical communication constraints, sensors are connected to other sensors located in their spatial neighborhood. Yet, we already noticed the degree was spatially distributed (nodes closer to the periphery have smaller degrees). Therefore, sensors tend to be connected to sensors with similar degrees. Fig. 2-c displays $\langle k_{nn} \rangle_k$ for the time step corresponding to the minimal correlation ($\rho = 0.1$). This slightly negative value can be explained by the relatively high average neighbor degree observed for the least connected nodes.

## 5. Distance-based measures

**Classic distance.** The (directed) *distance* between two nodes corresponds to the length of the shortest (directed) path between them. Authors studying the Internet are generally interested in this quantity, especially at the routing level, because it corresponds to the length of the optimal route between the considered nodes. The *diameter* of a network is the longest shortest path over all couples of nodes. It has been popular when studying the Internet [2, 4], because it corresponds to the length of the worst optimal path, and can therefore be interpreted as a measure of how well connected the network is. The *average distance* $\ell$ is also studied, as a summary of the distance distribution [3-5, 10]. When $\ell$ grows logarithmically relatively to $n$, then the network is said to be *small-world*, because any node can reach any other node in a few hops. The *distance distribution* itself is more informative, and is therefore also discussed [5, 10]. Sometimes authors do not study directly this distribution, but a related notion such as the *hop plot*. It originally corresponds to the total number of pairs of nodes separated by a certain distance (or closer), expressed as a function of this distance [4], but variants exist [3, 10].

For real-world systems, the diameter generally increases because the network grows. Interestingly, in our simulation, it is on the contrary because it shrinks. Changes in the diameter are likely to correspond to important modifications in the WSN structure, allowing to discover key moments in the simulation. These can be compared to those already identified by using other measures such as the maximal degree (Fig. 2-c). The average distance also increases with time, which means nodes get farther and farther. The distance distribution, considered just before and after diameter changes, confirms the appearance of longer shortest paths.

**Sink-distance.** In the case of WSNs though, the end-to-end communication does not take place between any two nodes, but only from one sensor to one sink. For this reason, some distance-based notions defined in the previous paragraph, such as the eccentricity, are not relevant here. To solve this, we define the concept of *sink-distance*: the distance between a sensor and the closest sink in the WSN. We call *sink-radius* the maximal sink-distance over all sensors. It is characteristic of the WSN since it represents how far a sensor can get from its closest sink. The term radius refers here to the fact the sinks hold a central position in the WSN, if not spatially at least functionally.

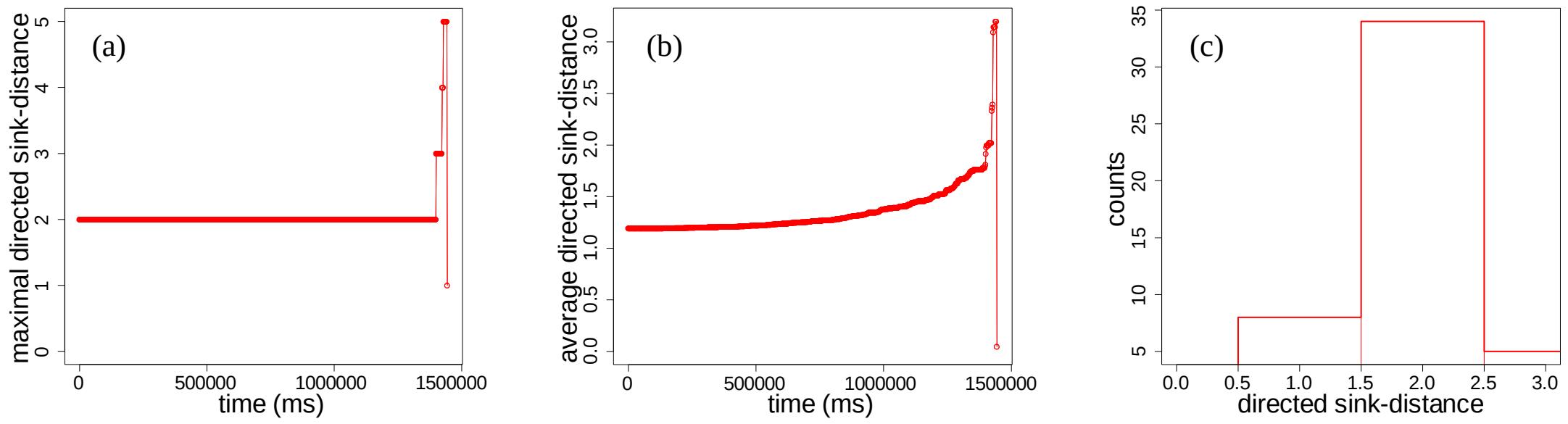

Fig. 3. (a) directed sink-radius and (b) average sink-distance, in function of time; (c) directed sink-distance distribution at $t = 1402200$.

During our simulation, the average sink-distance evolves much like the classic average distance (Fig. 3-b). The changes observed for the sink-radius (Fig. 3-a), however, do not temporally match those of the diameter. This highlights the fact the diameter reflects events which have little importance for a WSN. Indeed, an increase in the diameter corresponds to an increase in the distance between two sensors. Unless one belongs to the other's route toward the sink, there is no reason for this change to affect their communication with the sinks. Fig. 3-b displays the sink-distance at $t = 1402200$, showing most sensors need to go through one other sensor to reach the sink, while only a few of them can access it either directly or through two other sensors.

## 6. Centrality measures

**Classic betweenness.** Node centrality measures allow quantifying the importance of a node relatively to certain criteria. Moreover, by considering how such a measure behaves for all nodes, it is possible to characterize the whole network. Various measures have been used to study the Internet [4], but the most popular is the *betweenness* $b$ defined by Freeman [12]. In its original form, it corresponds to the number of shortest paths going through the node of interest, when considering all other pairs of nodes in the network. Freeman proposed to normalize the resulting value by dividing it by the theoretical maximum, in order to get a measure ranging from 0 to 1. However, in the domain of computer networks, most authors simply divide it by $n$ [10]. In this domain, the interpretation of the betweenness is straightforward: if we suppose the optimal path between two nodes is the shortest path, then the betweenness of a node corresponds to the amount of traffic it relays [5]. For this reason, the betweenness is sometimes called *load* in this context. It is studied under various forms: average betweenness $\langle b \rangle$ [3, 5], betweenness distribution [3, 5], average betweenness as a function of the degree $\langle b \rangle_k$ [3, 10], average neighbor betweenness as a function of the node betweenness $\langle b_{nn} \rangle_b$ [3].

In our experiment, the network is shrinking with time, so the number of shortest paths decreases, making in turn the betweenness decrease too. For this reason, we considered the normalized version of the betweenness (Freeman's version), in order to get comparable values. This explains the very low values observed in the following figures, which is not a problem considering we want to compare them. Another factor for this low values is the initial density of the network, which is very high, and stays well above the level of other real-world systems even at the end of the simulation (Fig. 2-b).

The average betweenness in the connected component increases regularly with time. This means the sensors remaining connected to the sink become more and more central relatively to the connected component size. The betweenness is a monotonic increasing function of the degree, as what was already observed for the Internet [10]. In other words: the higher the degree and the higher the centrality. This is true from the beginning to the end of the simulation. On the contrary, unlike the Internet [3], the average neighbor betweenness seems to be independent from the betweenness of the node of interest. In other words, the centrality of a node does not affect its neighbors' centrality.

**Sink-betweenness.** The betweenness is very interesting for WSNs, because the amount of traffic a node has to relay is directly related to its lifetime, which in turn has a direct effect on the network lifetime, a critical aspect in WSNs research. However, as we stated previously, in a WSN the traffic goes from a sensor to a sink. So here again, it seems more appropriate to consider only this kind of paths. On the model of the classic betweenness, we define the *sink-betweenness* of a node as the proportion of sensor-to-sink shortest paths going through it.

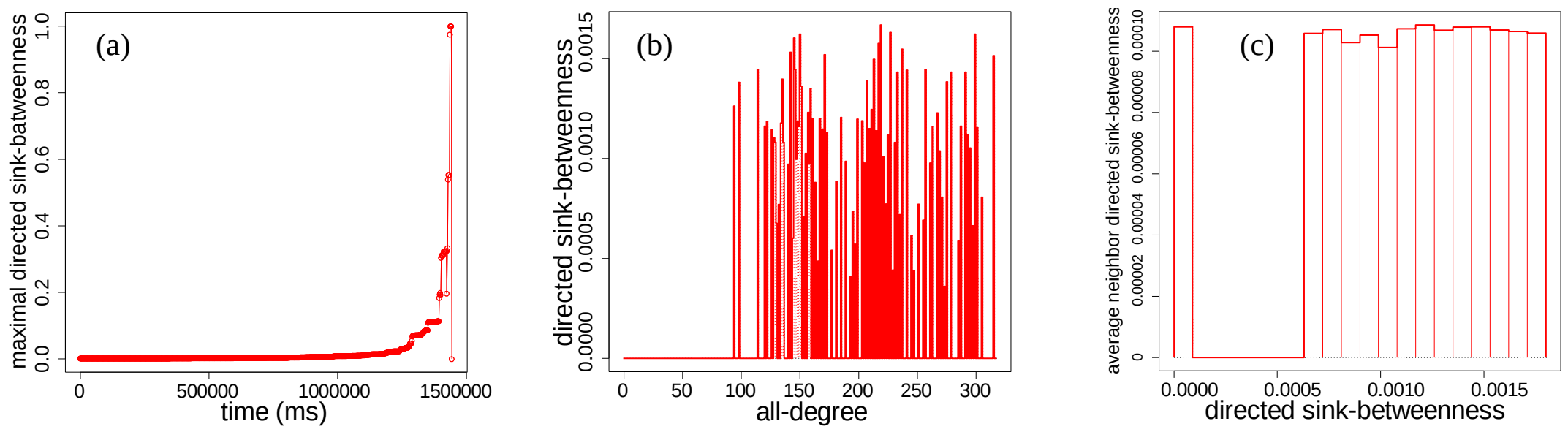

Fig. 4. (a) maximal directed sink-betweenness in function of time; (b) average directed sink-betweenness in function of the degree and (c) in function of the sink-betweenness, at $t = 0$.

Like the classic betweenness, and for the same reasons, the measured sink-betweenness is initially very low. The evolution of the average sink-betweenness is also very similar to what was observed for the classic betweenness. A more interesting measure is the maximal sink-betweenness, represented in Fig. 4-a, because it allows detecting the presence of particularly central sensors. It is unlikely to observe those at the beginning of the simulation, due to the network density. Their appearance therefore corresponds to critical steps in the evolution of the network. In our simulation, the maximal sink-betweenness increases strongly at the end of the simulation, reaching 1 when only a single node is directly connected to the sink, i.e. when all communication must go through it. Sensors with a high sink-betweenness are likely to consume their energy rapidly, because of the high number of requests they receive from other sensors. Fig. 4-b shows the sink-betweenness expressed as a function of the degree, at the beginning of the simulation. Unlike the classic betweenness, there is no clear relationship, which shows the measures behave differently. On the contrary, the average sink-betweenness of a node neighbors seems to be independent from its own sink-betweenness Fig. 4-c, as for the classic betweenness.

## 7. Conclusion

In this article, we described the different tools one can use to analyze the topology of WSNs. Our contribution is three-fold. First, we performed a review of the most popular measures used in existing studies regarding the Internet. We discussed their appropriateness to the study of WSNs, and pointed out their limitations in this context. Second, we defined new, WSN-specific measures aiming at solving these limitations, and explained how they could be used. Finally, we simulated some WSN activity thanks to Opnet in order to illustrate the relevance of both retained classic measures and newly proposed ones. Taking advantage of this data, we explained how their values can be interpreted to characterize the evolution of a WSN. We focused on five classes of measures. For those relative to the *network size* (node and link counts, density), we proposed to distinguish the live part of the WSN from the disconnected sensors. We showed how one could use the existing *degree-based* measures (average, maximal, minimal degree, degree distribution, average neighbor degree, etc.) to identify critical points in the evolution of WSNs, which are characterized by a very dynamic nature. We introduced the notion of sink-distance, in order to adapt the existing *distance-based* measures (diameter, average distance, distance distribution, etc.) to WSNs. We defined the related notion of sink-betweenness in order to extend the *centrality-based* measures traditionally used to study the Internet (average betweenness, betweenness distribution, etc.). Finally, we applied *transitivity-based* measures to highlight a very uncommon trait of the studied WSN: its transitivity stays constant despite the fact it never stops losing nodes and links during the simulation.

This work can be extended in several ways. First, for space matters, we could not be exhaustive regarding the reviewed topological properties. For instance, other centrality measures exist and are used to characterize computer networks [4]. More importantly, we had to ignore the mesoscopic structure of the network, which is nevertheless of great interest. It can be studied by considering the presence of strongly connected components (i.e. maximal subgraphs in which all nodes are interconnected through directed paths) or communities (subgraphs more densely interconnected relatively to the rest of the network). Finally, we did not consider the transitivity (a.k.a. clustering coefficient), which is also a very important topological measure. Another interesting aspect concerns the measures we defined. As we stated before, although informative, traditional measures are not always relevant to study WSNs, due to their specific features, which is why we introduced our own ones. Consequently, these could be applied to networks representing different real-world systems, provided those are similar enough to WSNs. Let us consider, for instance, an egocentric social network. It is extracted by selecting first a person of interest as the center of the network, and adding its direct and indirect acquaintances, until some radius limit is reached. In this context, the sink-distance and sink-betweenness would be of great interest to characterize acquaintances relatively to the person of interest.

## Acknowledgements

This work is supported by the Galatasaray University Research Foundation under the Grant No. 10.401.006.